\documentclass[aps,twocolumn,superscriptaddress]{revtex4-1}
\usepackage{graphicx}
\usepackage{bm}
\usepackage{braket}
\usepackage{pifont}
\usepackage{amsmath}
\usepackage{amsfonts}
\usepackage{amssymb}    
\usepackage{graphicx}
\usepackage{color,xcolor}
\usepackage{tensor}
\usepackage[caption=false]{subfig}
\usepackage{natbib}
\usepackage{dsfont}
\usepackage{hyperref}
\hypersetup{
    colorlinks=true,
    linkcolor=blue,
    filecolor=black,   
    citecolor=blue, 
    urlcolor=blue,
   	pdftitle={Quantum repeater for continuous variable entanglement distribution},
    bookmarks=true,
}
\newcommand {\mrm}[1] {\mathrm{#1}}

\begin{document}
\title{Quantum repeater for continuous variable entanglement distribution}
\author{Josephine Dias}
\email{jdias@nii.ac.jp}
\affiliation{National Institute of Informatics$,$ 2-1-2 Hitotsubashi$,$ Chiyoda$,$ Tokyo 101-0003$,$ Japan.}
\author{Matthew S. Winnel}
\affiliation{Centre for Quantum Computation and Communication Technology$,$ School of Mathematics and Physics$,$ University of Queensland$,$ Brisbane$,$ Queensland 4072$,$ Australia.} 
\author{Nedasadat Hosseinidehaj}
\affiliation{Centre for Quantum Computation and Communication Technology$,$ School of Mathematics and Physics$,$ University of Queensland$,$ Brisbane$,$ Queensland 4072$,$ Australia.} 
\author{Timothy C. Ralph}
\affiliation{Centre for Quantum Computation and Communication Technology$,$ School of Mathematics and Physics$,$ University of Queensland$,$ Brisbane$,$ Queensland 4072$,$ Australia.} 
\date{\today}

\begin{abstract}
Quantum repeaters have been proposed as a way of extending the reach of quantum communication. First generation approaches use entanglement swapping to connect entangled links along a long distance channel. Recently, there have been proposals for first generation quantum repeaters for continuous variables. In this paper, we present an improved continuous variable repeater scheme using optimal Gaussian entanglement swapping. Our scheme uses the noiseless linear amplifier for entanglement distillation. We show that with the simplest configuration of the noiseless linear amplifier and under the assumption of good quantum memories and perfect sources and detectors, our scheme beats the direct transmission upper limit for shorter distances and can offer advantages over previous CV repeater schemes.
\end{abstract}
\maketitle
\section{Introduction}
The development of technologies according to the principles of quantum mechanics allows many promising real world applications. Under the umbrella term of quantum communication \cite{gisin2007quantum}, these applications range from secure communication 
\cite{gisin2002quantum, scarani2009security, pirandola2019advances} and quantum state transfer \cite{pirandola2015advances}, to enhanced quantum sensing \cite{ge2018distributed,proctor2018multiparameter, zhuang2018distributed} and computation \cite{grover1997quantum,cirac1999distributed}. However, utilizing these technologies over long distances remains challenging due to fiber loss or free space attenuation. In classical communication, this problem is solved by having repeaters stationed at various points along the channel to amplify the signal. This solution, that has enabled classical communication to proceed, may not be employed for quantum communication as redundant copies of quantum information cannot be made due to the no-cloning theorem \cite{wootters1982single}. A more sophisticated solution is necessary if these issues are to be overcome and we are able to utilize the advantages of quantum communication over long distances. 

One proposed solution has come in the form of a quantum repeater \cite{briegel1998quantum}. The first quantum repeater protocol from the late nineties used multiple rounds of entanglement swapping \cite{ziukowski1993event} in order to connect entangled pairs, and share entanglement between ends of a long distance channel. Entanglement purification \cite{bennett1996purification} was also required to correct against building operation errors. Since this first proposal, there has been significant theoretical advancement on repeater protocols and experimental progress with repeater elements \cite{munro2015inside,muralidharan2016optimal, sangouard2011quantum}. 

Currently, the majority of repeater proposals are for discrete variable (DV) encodings of quantum information \cite{sangouard2011quantum, munro2015inside, muralidharan2016optimal}, where information is encoded in a finite dimensional basis, such as the polarization of single photons. As an alternative, there are also continuous variable (CV) encodings of quantum information, where information is encoded in the quadrature amplitudes. Not only do continuous variable encodings of quantum information offer (in principle) easier state generation, manipulation and detection \cite{braunstein2005quantum}, they also offer the possibility of compatibility with existing infrastructure \cite{kumar2015coexistence}.   

In the past few years, there have been three different proposals for the first generation of continuous variable quantum repeaters \cite{dias2017quantum, furrer2018repeaters, seshadreesan2020continuous}. 
In this paper, we present an improvement upon one of these previous CV quantum repeaters, the protocol presented in Ref.~\cite{dias2017quantum}. Like Refs.~\cite{dias2017quantum, seshadreesan2020continuous}, our repeater uses the single quantum scissor (QS) to distill CV entangled states. Unlike Ref.~\cite{seshadreesan2020continuous}, which uses non-deterministic non-Gaussian entanglement swapping, our CV quantum repeater uses  Gaussian entanglement swapping with post-selection. Our scheme utilizes a different Gaussian entanglement swapping setup than Ref.~\cite{dias2017quantum}, thus we are able to report an improvement in the attainable key rates. The CV repeater scheme presented in this paper surpasses a fundamental upper limit on quantum communication via direct transmission (the so-called PLOB bound) \cite{pirandola2017fundamental} for a total distance of 322 km.

We emphasize that the results in this paper do not model the effects of imperfect quantum memories, sources or detectors. While it can be expected that incorporating these effects into our results would inevitably decrease the key rates, the preliminary results reported here represent a useful step towards distributing CV entanglement and implementing CVQKD over long distances. This paper is arranged in the following way: in Sec.~\ref{sec:repeater} we explain the structure of our CV repeater and in Sec.~\ref{sec:results} we will present results. Finally, in Sec.~\ref{sec:conc} we provide some future directions based on our findings and conclude.
\section{CV quantum repeater \label{sec:repeater}}
First generation quantum repeaters are based on three core elements: entanglement distribution, entanglement swapping, and entanglement purification (or distillation) protocols (see Ref.~\cite{munro2015inside} for a review). In the following section, we will give an overview of how each of these elements will be implemented in our repeater. 
\begin{figure}[h!]
\centering
\subfloat[]{%
  \includegraphics[width=0.99\linewidth]{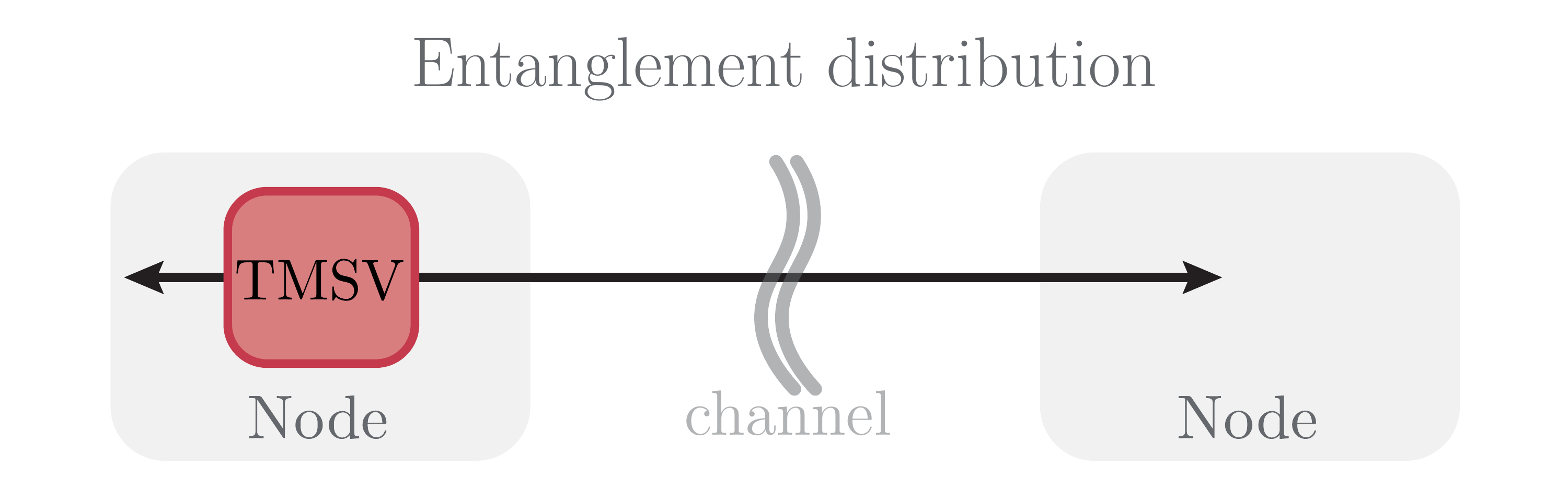} \label{fig:entdist}%
}

\subfloat[]{%
  \includegraphics[width=0.99\linewidth]{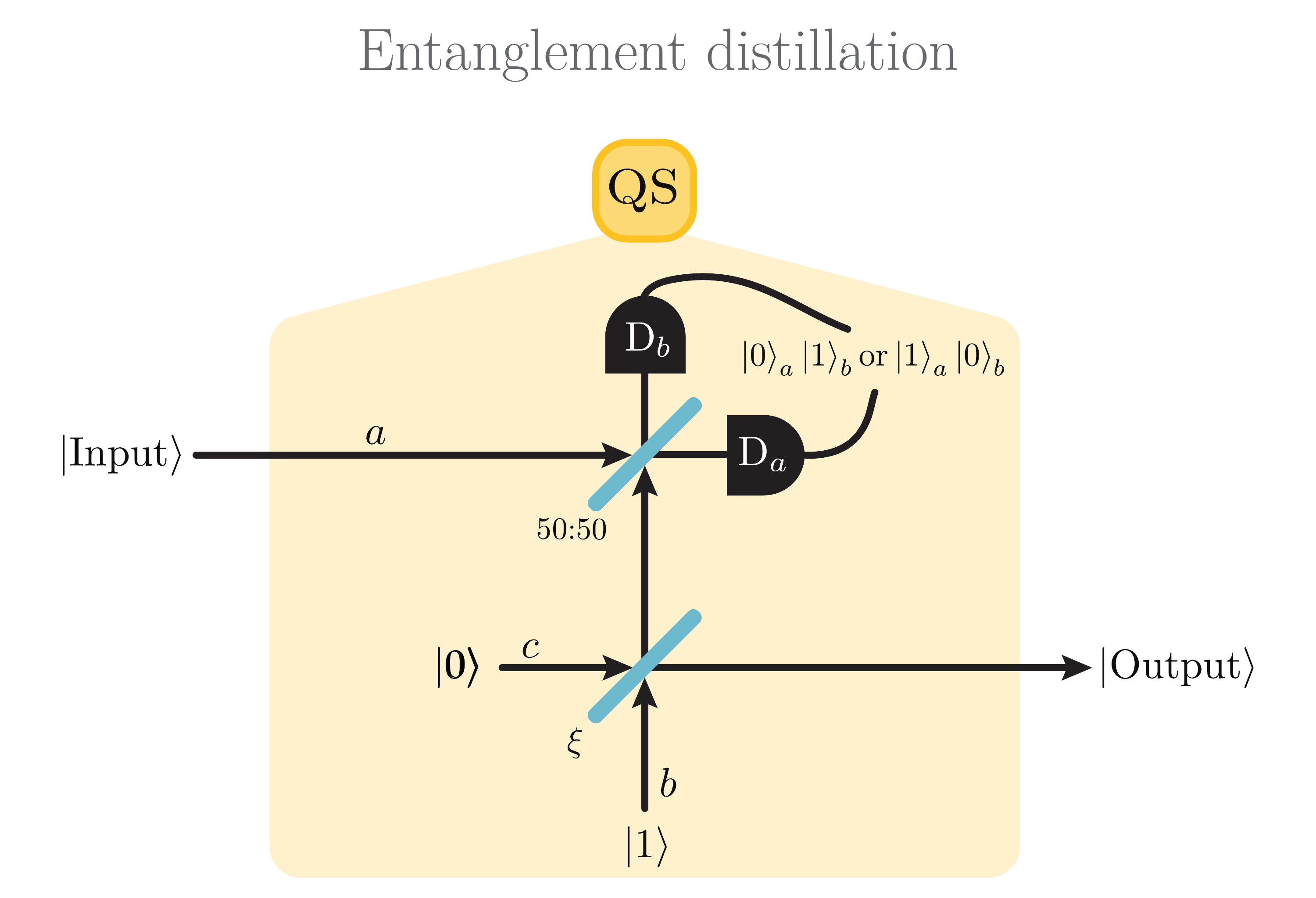}
  \label{fig:NLA}
  }

\subfloat[]{%
  \includegraphics[width=0.99\linewidth]{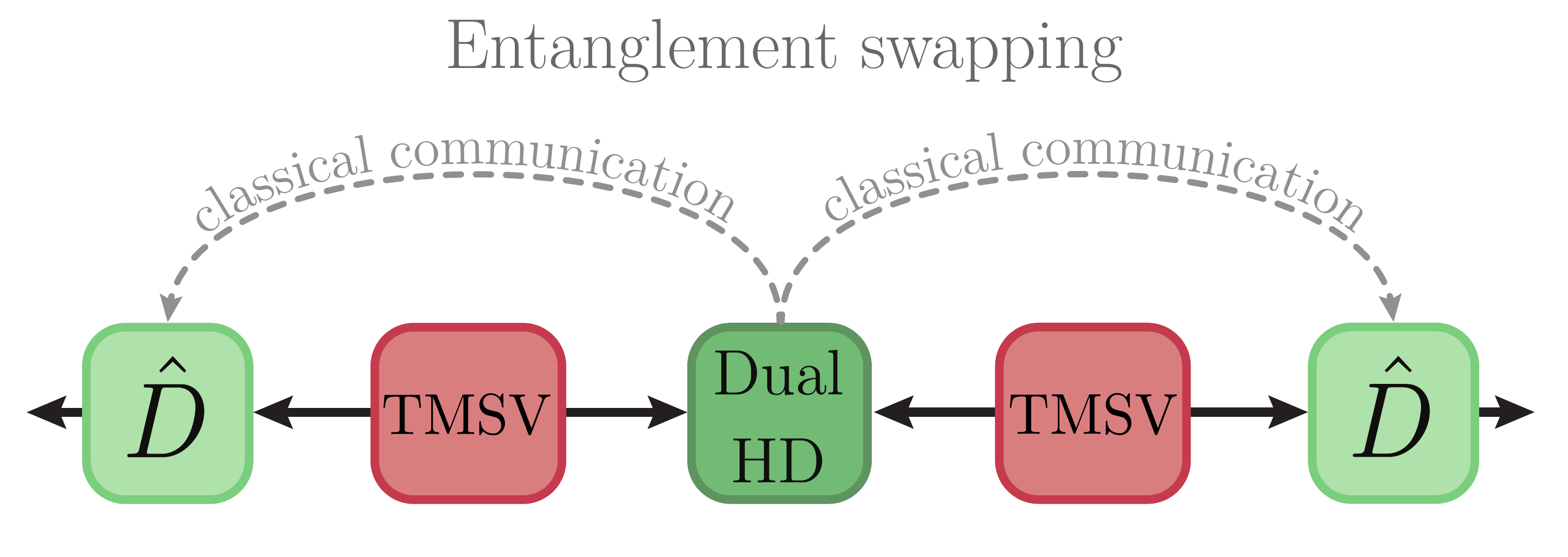}
  \label{fig:entswap1}
  }
\caption{Components of the CV quantum repeater \protect\subref{fig:entdist} Entanglement distribution in the CV quantum repeater. One mode of a TMSV state is sent through a lossy channel to a neighboring repeater node. The other mode of the entangled state remains in the same node. \protect\subref{fig:NLA} The CV repeater uses the NLA to distill entangled TMSV states. The simplest linear optics construction of the NLA is pictured here consisting of a single quantum scissor (QS). The input is combined with an ancilla photon which has passed through a beam-splitter of tunable ratio \(\xi\), this is related to the gain of the NLA via \(g=\sqrt{(1-\xi)/\xi} \). The combined modes are detected and success is heralded when a single photon is detected at one output and none at the other. \protect\subref{fig:entswap1} Gaussian entanglement swapping protocol from Ref.~\cite{hoelscherobermaier2011optimal}. Modes of two independent TMSV states are combined and input into a dual homodyne detection (dual HD). The results of the detection are sent in both directions to both output modes where displacements are performed accordingly.}
\label{fig:Components}
\end{figure}
\subsection{Entanglement distribution}
Beginning with entanglement distribution, the entangled resource states used in our protocol are the Gaussian two-mode squeezed vacuum (TMSV) state:
\begin{equation}
\ket{\chi}_{ab} = \sqrt{1-\chi^2} \sum \chi^n \ket{n}_{a} \ket{n}_{b} .
\label{eq:chi5}
\end{equation}
where \(0<\chi<1\) is the two-mode squeezing parameter.
Distribution of these states \eqref{eq:chi5}, is performed asymmetrically (see Fig.~\ref{fig:entdist}) with entangled states being generated at each node of the quantum repeater and then one mode of the entangled state is passed through a lossy channel through to the neighboring node. One mode of each of these entangled states would be decohered by loss from transmission through the channel while the other mode remains untouched in the same node.

\subsection{Entanglement distillation}
In our CV repeater, entanglement distribution in the repeater links is followed immediately by distillation on the entangled mode that has passed through the lossy channel. Entanglement distillation is a necessary component in first generation repeaters, needed to combat the  decoherence effects from channel loss and entanglement swapping operations. In the scheme of Refs.~\cite{dias2017quantum, seshadreesan2020continuous} and in this work, the Noiseless Linear Amplifier (NLA) \cite{ralph2009nondeterministic} is used to distill the entangled states. When implemented with linear optics, the simplest NLA comprises of a single modified QS device \cite{ralph2009nondeterministic}. The single QS implementation of the NLA has been demonstrated experimentally \cite{kocsis2012heralded,xiang2010heralded}, and more specifically entanglement distillation on TMSV states decohered by loss has been demonstrated with a similar device \cite{ulanov2015undoing}.

\subsection{Entanglement swapping}
Following entanglement distribution and distillation, our CV repeater will use deterministic Gaussian entanglement swapping \cite{loock1999unconditional} in order to connect the entangled repeater links. We employ the optimal Gaussian entanglement swapping protocol described in Ref.~\cite{hoelscherobermaier2011optimal}. This involves sending classical signals to both ends of the channel and conducting displacements on both modes (see Fig.~\ref{fig:entswap1}). This is unlike other protocols (including CV teleportation) where classical communication and displacements are only performed on one mode. In this way, two pure Gaussian entangled states can be swapped and the resulting entangled state remains pure. In general, for any two Gaussian states, entanglement swapping in this way is optimal  \cite{hoelscherobermaier2011optimal}. 

The use of the optimal Gaussian entanglement swapping scheme of Ref.~\cite{hoelscherobermaier2011optimal} represents the main difference between this work and the work in Ref.~\cite{dias2017quantum} which used CV teleportation. In this work we also consider the use of post-selection based on the results of the dual HD in the swapping scheme. Qualitatively, this means that based on the results of the dual-HD some results will be rejected and some will be accepted, thus entanglement swapping in our repeater is not deterministic. Post-selection in our scheme is necessary because the truncation due to the single quantum scissor deteriorates the raw key and adds non-Gaussianity. This effect is more pronounced for large measurement outcomes and thus we use post-selection to filter the measurement results, accepting results that are close to 0. 

\section{Results \label{sec:results}}
\subsection{Single node repeater}
\begin{figure}
\centering
  \includegraphics[width=0.99\linewidth]{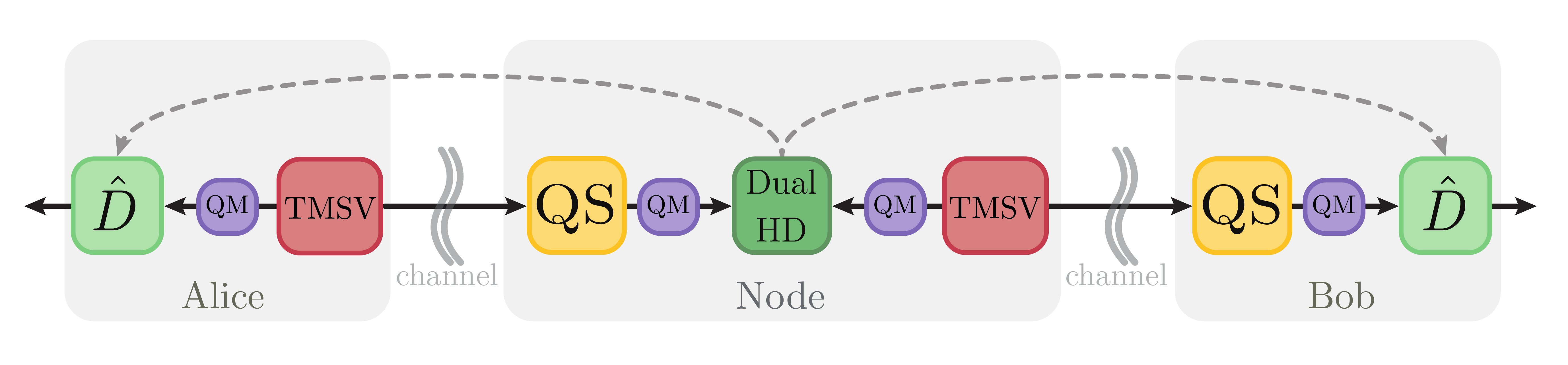}
\caption{Simplest implementation of an improved first generation CV repeater. The entangled resource states used are TMSV states (given by \eqref{eq:chi5}). Entanglement distillation is performed by a QS. Once successful distillation has been heralded, the distilled state is stored in a quantum memory (QM) where it will wait for the neighboring quantum scissor to succeed. Gaussian entanglement swapping is conducted by a dual HD of the two modes at the repeater node, if the outcome of the dual HD is within the accepted post-selection range around zero, this is followed by classical communication of the results of the detection being sent to Alice and Bob and both modes are then displaced accordingly (\(\hat{D}\)). This configuration is asymmetric as the two inputs to the dual HD are not the same. This setup requires one source to be placed with Alice and a quantum scissor to be placed at the repeater node.}
\label{fig:asymrep1}
\end{figure}
The simplest implementation of our improved CV repeater protocol combing all the aforementioned elements, is shown in Fig.~\ref{fig:asymrep1}. It is formed using a single repeater node in the center of the channel with NLAs implemented in their simplest configuration (consisting of a single quantum scissor). Entanglement distribution is performed by sending one mode of a TMSV state \eqref{eq:chi5} through the channel between the single repeater node and ends of the channel. The mode of the entangled state that had passed through the lossy channel is then distilled using the single quantum scissor. 

While the quantum scissor operation is non-deterministic, both entangled states are independent at this point in the protocol, therefore both quantum scissors can operate independently and simultaneously. When a quantum scissor heralds successful operation, we assume high quality quantum memories are available to store the distilled entanglement until the other quantum scissor is successful. After both entangled states have been distilled, they are then swapped by mixing the two modes at the repeater node and conducting a dual homodyne detection. For the results to be accepted, the measurement outcome of the dual HD results must fall within a certain radius around zero (see Appendix~\ref{app:single} for details). If this is successful, the results of this detection are then sent to Alice and Bob and a displacement is performed on each mode based on the results of the detection which completes the entanglement swapping operation.   
\begin{figure}
\centering
\includegraphics[width=0.99\linewidth]{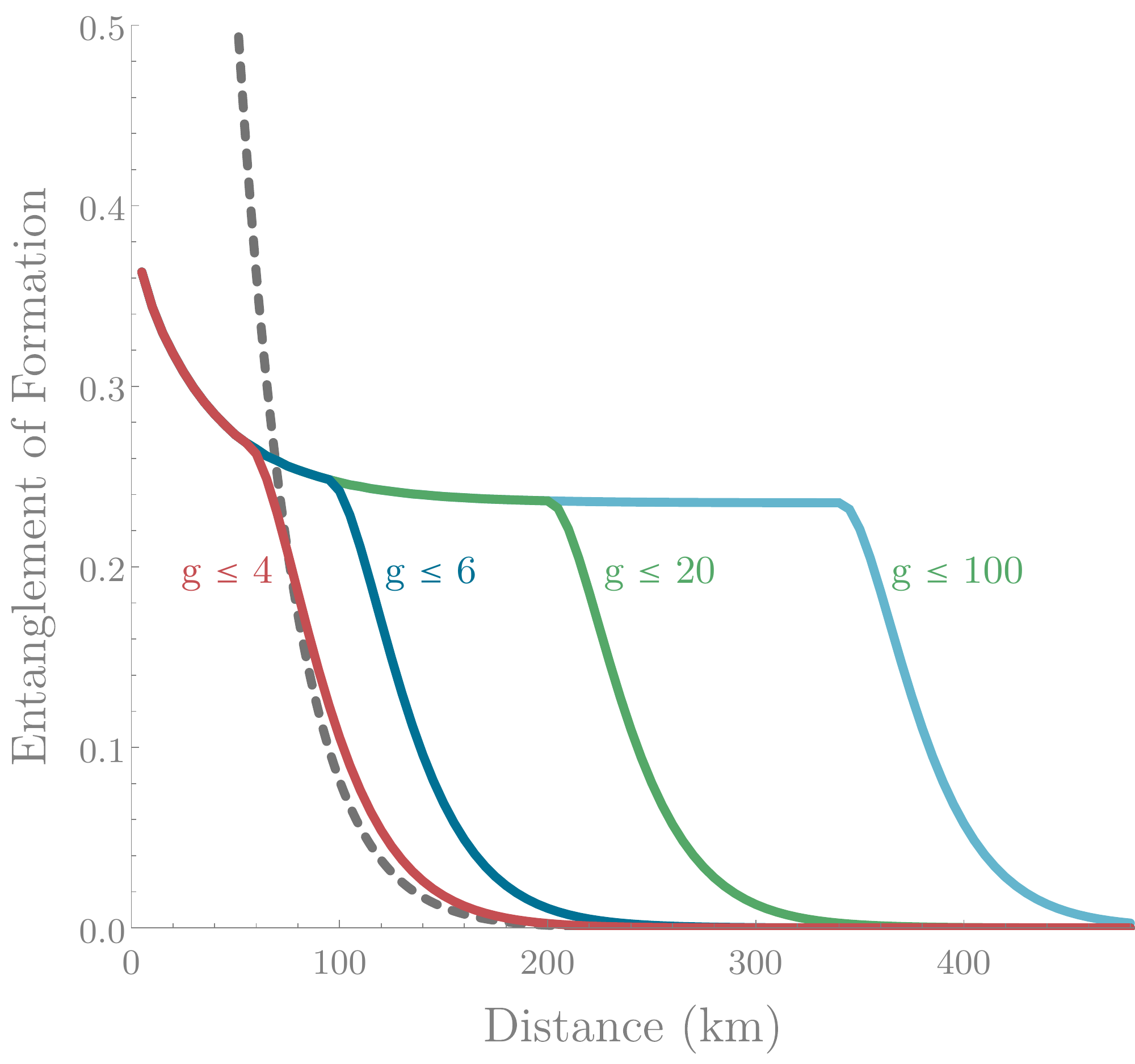}
\caption{Entanglement of formation of the CV repeater. The solid, colored lines show the entanglement of formation between ends of the channel using the single-node CV repeater with TMSV state sources of squeezing \(\chi=0.3\). Each line shows the optimal EOF attainable when the NLA gain has been restricted to some maximum value (each line has been labelled with this maximum gain). The dashed, dark gray line is the entanglement of formation of an unphysical, infinitely squeezed TMSV state (Eq.~\eqref{eq:chi5} with \(\chi\to1\)) transmitted through an optical fibre channel of the same distance. }
\label{fig:eof}
\end{figure}

Initially, we consider the maximum entanglement that can be distributed via our repeater by evaluating the entanglement of formation (EOF) \cite{wolf2004gaussian,akbarikourbolagh2015entanglement,marian2008entanglement} between the end stations when post-selection of the HD results lying very close to zero are accepted. This result is given in Fig.~\ref{fig:eof} where we show the entanglement of formation between end stations of our CV repeater using TMSV sources of fixed squeezing \(\chi=0.3\). We compare this to the EOF of an unphysical, infinitely squeezed TMSV state distributed through the same loss. Each solid line in Fig.~\ref{fig:eof} shows the highest EOF achievable for various maximum NLA gains. At shorter distances, EOF maximises for lower gains. It can be seen on Fig.~\ref{fig:eof} that there is a turning point on each solid line. This turning point marks the distance beyond which maximum EOF is achieved by the maximum allowable NLA gain. For maximum gains of 5 or higher, the EOF surpasses the direct transmission EOF at a distance of 70km.  While the red line for \(g\leq4\) does produce an improvement over the direct transmission EOF, this occurs for a total channel distance of 75km. Additionally, for NLA gains of \(g\leq 3\) we do not observe an improvement beyond the direct transmission EOF. Note that in Fig.~\ref{fig:eof} and throughout this paper we have considered optical fiber with loss of 0.2 dB/km.

\begin{figure}
\centering
\includegraphics[width=0.99\linewidth]{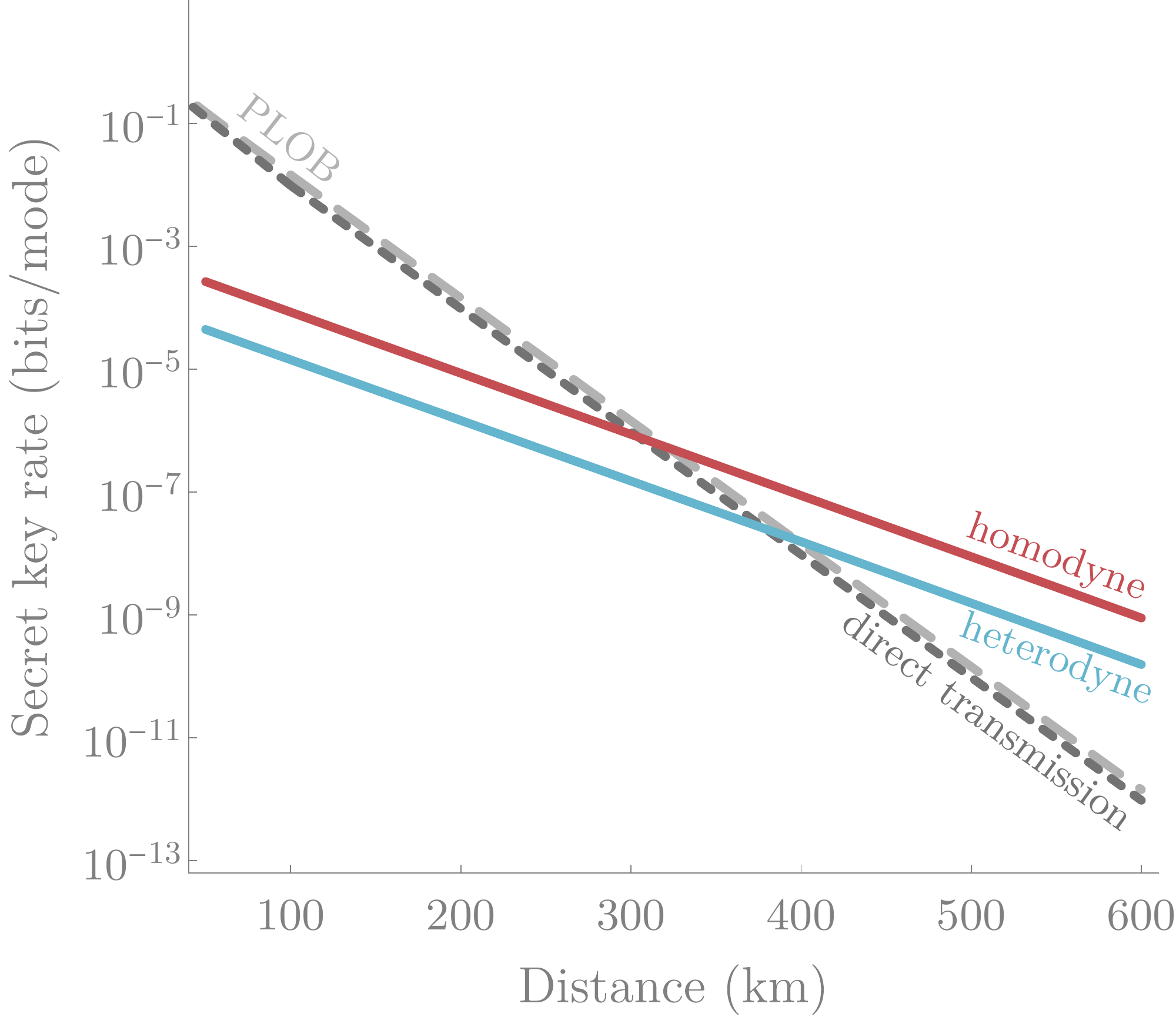}
\caption{Secret key rate of the single node repeater shown in Fig.~\ref{fig:asymrep1} using single quantum scissors. The blue line shows the key rate for a heterodyne-based QKD protocol and post-selection cut-off at \(\gamma_{\mrm{max}}=0.4\).  The red line shows the same result except with a homodyne-based QKD protocol and a larger post-selection cut-off of \(\gamma_{\mrm{max}}=0.5\). The dashed, dark gray line is the key rate (using a homodyne-based CVQKD protocol) for direct transmission through the channel  using optimized, finite squeezing (\(\chi<1\)). Reconciliation efficiency for the homodyne, heterodyne and direct transmission lines has been set to \(95\%\).  The dashed, light grey line is a fundamental upper bound on the maximum secret key rate that is achievable using direct transmission without a quantum repeater (PLOB bound) \cite{pirandola2017fundamental}. }
\label{fig:result1}
\end{figure}

As a second figure of merit, which importantly incorporates the probability of success, we consider the secret key rate achievable by the CV quantum repeater. We use the secret key rate as it ensures the results  are comparable with previous CV repeater proposals \cite{furrer2018repeaters, dias2020quantum}. The secret key rate of the scheme shown in Fig.~\ref{fig:asymrep1} is able to surpass the absolute maximum secret key rate for direct transmission (shown by the dashed, grey line on Fig.~\ref{fig:result1} and referred to as the PLOB bound \cite{pirandola2017fundamental}). The secret key rate presented in Fig.~\ref{fig:result1} is defined as:
\begin{equation}
\mathrm{Secret\, key\, rate} = K \times R_{\mrm{rep}}
\end{equation} 
where \(K\) is the raw key rate calculated from the covariance matrix of the output state (see details in Appendix~\ref{app:single}), and \(R_{\mrm{rep}}\) is the rate of successful operation of the entire repeater which depends on the success probability of the QS and post-selection. The rate \(R_{\mrm{rep}}\) for successful operation of the CV repeater for \(2^n\) links is calculated via
\begin{equation}
R_{\mrm{rep}} = \frac{1}{Z_n\left(P_{\mrm{NLA}}\right) }\times   \prod_{i=0}^{n-1} \frac{1}{Z_i\left(P_{\mrm{PS}i}\right) } 
\label{eq:reprate}
\end{equation}
where \(P_{\mrm{PS}i} \) is the probability of successful post-selection in the various entanglement swapping rounds when \(2^i\) swaps need to occur successfully for the repeater protocol to proceed. Here, \(i=n-1\) corresponds to the first round of entanglement swapping with  \(2^{n-1}\) swaps, and \(i=0\) corresponds to the final swap. The function \(Z_n\left(P\right)\) is the average number of steps to generate successful outcomes in \(2^n\) probabilistic operations, each with success probability \(P\) \cite{bernardes2011rate}:
\begin{equation}
Z_n \left(P\right) =\sum^{2^n}_{j=1} \binom{2^n}{j} \frac{\left(-1\right)^{j+1}}{1-\left(1-P\right)^j}
\end{equation}
For the single node results in Fig.~\ref{fig:result1}, the repeater rate \eqref{eq:reprate} is simply:
\begin{equation}
R_{\mrm{rep}}  =  \frac{1}{Z_1\left(P_{\mrm{NLA}}\right) } \times  \frac{1}{Z_0\left(P_{\mrm{PS0}}\right) }
\end{equation}

In Appendix~\ref{app:single}, we give details on how we calculate the entangled output state. Using this entangled state,  we calculated key rates for two entanglement-based CV-QKD protocols, one where Alice and Bob both perform heterodyne detection \cite{weedbrook2004quantum} and the other where they both perform homodyne detection \cite{cerf2001quantum} to their own entangled modes to obtain raw key.  These are shown on Fig.~\ref{fig:result1} by the blue and red lines respectively. The key rate shown is for reverse reconciliation, where Bob is the reference for reconciliation which is favorable in high loss regimes. Optimization of the normalized rate has been performed at each point over both gain of the QS's and strength of the TMSV state sources. Note that success probability of the QS decreases as gain is increased.  Optimal performance is achieved for squeezing of \(0.31< \chi_{\mathrm{opt}}<0.36\). 

In Fig.~\ref{fig:result1}, the keyrates for the homodyne based CVQKD protocol outperform the heterodyne protocol at all distances. This is because positive key can be achieved using the homodyne-based CVQKD protocol for larger post-selection cut-off regions. Larger cut-off regions correspond to a bigger post-selection success probability and thus increase the overall key rate. For the results in Fig.~\ref{fig:result1}, the heterodyne protocol uses a post-selection cut-off of \(\gamma_{\mrm{max}}=0.4\) which was found to be roughly optimal. However, the homodyne result uses a larger cut-off of \(\gamma_{\mrm{max}}=0.5\) and produces a higher key rate. 

Fig.~\ref{fig:result1} shows that the the PLOB bound is beaten for a total channel distance of 322km. The repeater is able to surapass the direct transmission key rate at 305km. This represents significant improvement upon single node operation of the CV  repeater in Ref.~\cite{dias2017quantum} which uses CV teleportation and beats the PLOB bound for distances above 500km \cite{dias2020quantum}. We emphasize these distances are total channel distances, meaning the point at which the protocol beats direct transmission, 305km, corresponds to 152.5km of optic fiber between Alice and the node (and between Bob and the node). 

\subsection{Multi-node repeater with nested swapping}
\begin{figure*}
\centering
\includegraphics[width=0.99\linewidth]{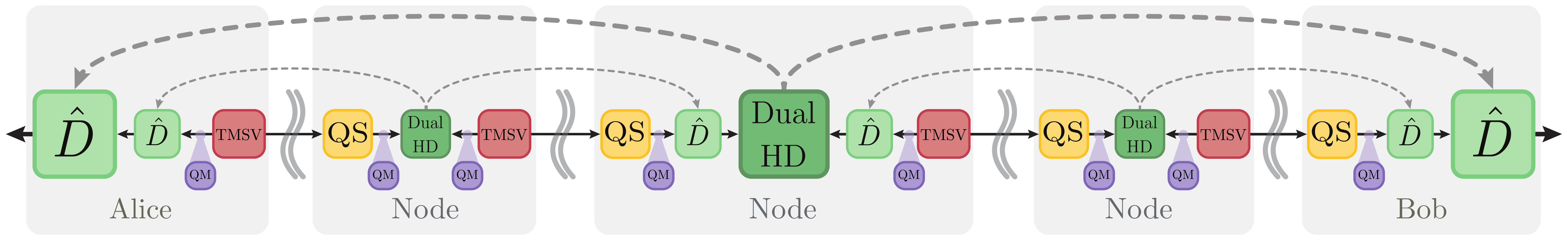}
\caption{Four links of the CV quantum repeater using one level of nested swapping. By comparison with Fig.~\ref{fig:asymrep1}, it can be seen that two independent implementations of the two link (single node) CV repeater are connected via nesting within another Gaussian entanglement swapping protocol (Fig.~\ref{fig:entswap1}). Quantum memories are required to hold the distilled entangled states until all quantum scissors are successful, then deterministic, nested entanglement swapping can proceed. }
\label{fig:nestedAsym}
\end{figure*}

\begin{figure}
\centering
\includegraphics[width=0.99\linewidth]{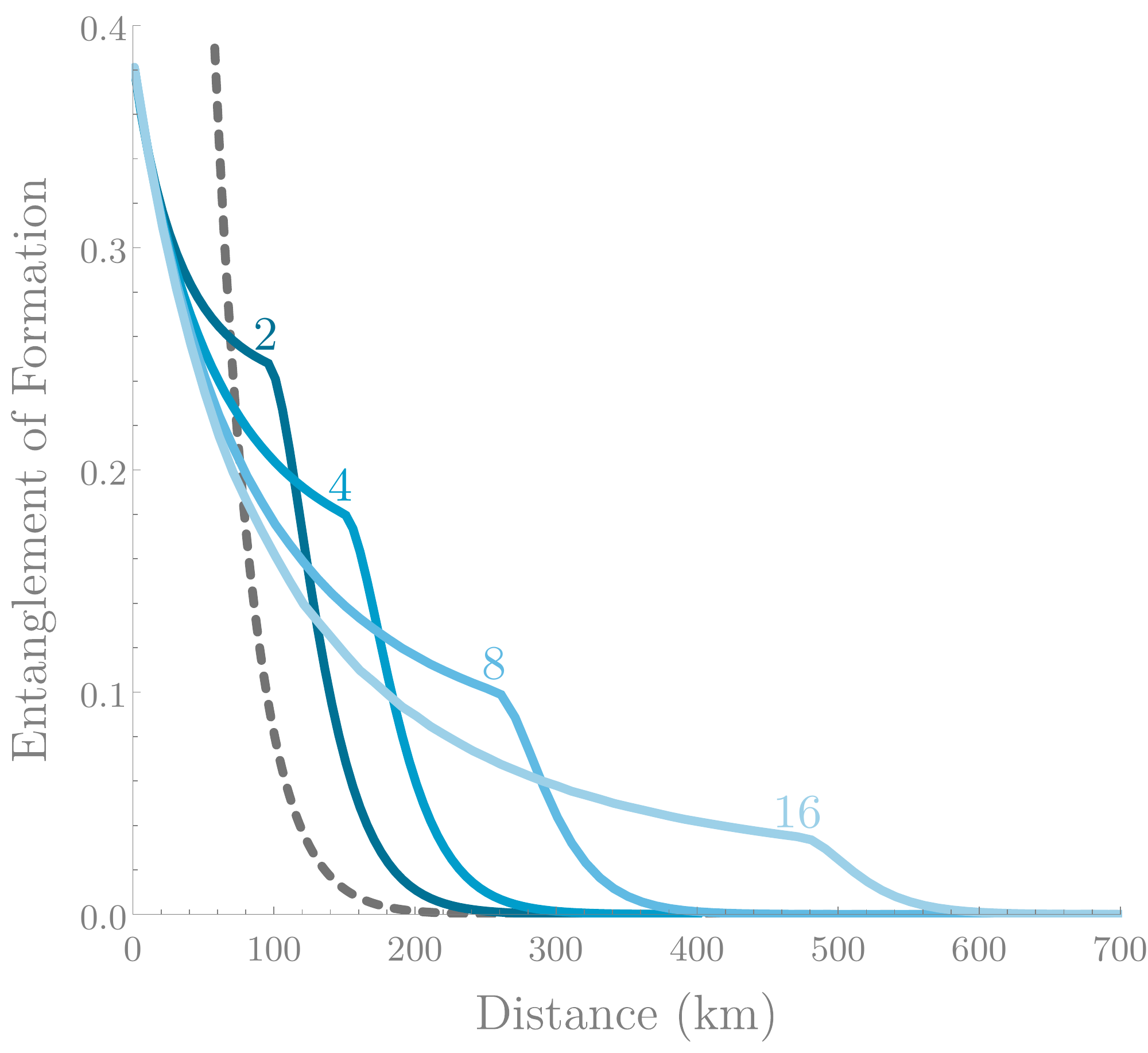}
\caption{Entanglement of formation of the CV quantum repeater for operation with two and more links. All solid blue lines show the EOF between end stations of the CV repeater for various numbers of repeater links (labelled). Like Fig.~\ref{fig:eof}, these results are achieved when post-selection of homodyne results lying close to 0 are accepted. The CV repeater uses an optimised NLA gain limited to \(g\leq 6\) and TMSV sources of squeezing \(\chi=0.3\). The dashed, dark gray line is the EOF of an infinitely squeezed TMSV state distributed through an optical fibre channel of the same total distance.}
\label{fig:eofAll}
\end{figure}

In order to use this CV quantum repeater over long distances, more nodes along the channel are required as well as more entanglement swapping operations to connect the entangled links.  To illustrate how this would proceed, see Fig.~\ref{fig:nestedAsym} with four links of the repeater connected via three repeater nodes. The protocol in Fig.~\ref{fig:nestedAsym} is just two copies of the asymmetric entanglement swapping protocol in Fig.~\ref{fig:asymrep1} connected via another Gaussian entanglement swapping with post-selection. 

For even longer distances and more repeater nodes, nesting proceeds in this way, where the output of two identical and independent copies of the protocol in Fig.~\ref{fig:nestedAsym} would be connected within another entanglement swapping operation. It is important to note that our repeater does not use nested entanglement distillation, meaning distillation occurs after entanglement distribution and not at any time after. Structuring the repeater in this way has an extremely favorable effect on the repeater rates, as it lowers the number of probabilistic operations occurring within the protocol. 

Again, we initially study the entanglement that may be distributed in this way. Like Fig.~\ref{fig:eof}, the results in Fig.~\ref{fig:eofAll} show the entanglement that may be distibuted between end stations of the CV repeater when  results lying very close to 0 are accepted. While we showed the effect of increasing gain on EOF in the results in Fig.~\ref{fig:eof}, for a fair multi-node comparison we restrict all NLA gains to the same maximum value in Fig.~\ref{fig:eofAll}; as an example, we use \(g\leq 6\). As expected, increasing the maximum NLA gain results in a larger distances that entanglement may be distributed. However, even with NLA gain restricted to \(g\leq 6\), Fig.~\ref{fig:eofAll} shows how our CV repeater may be used to distribute entanglement hundreds of kilometers beyond what is achievable using direct transmission with an unphysical, infinitely squeezed source.  

\begin{figure}
\centering
\includegraphics[width=0.99\linewidth]{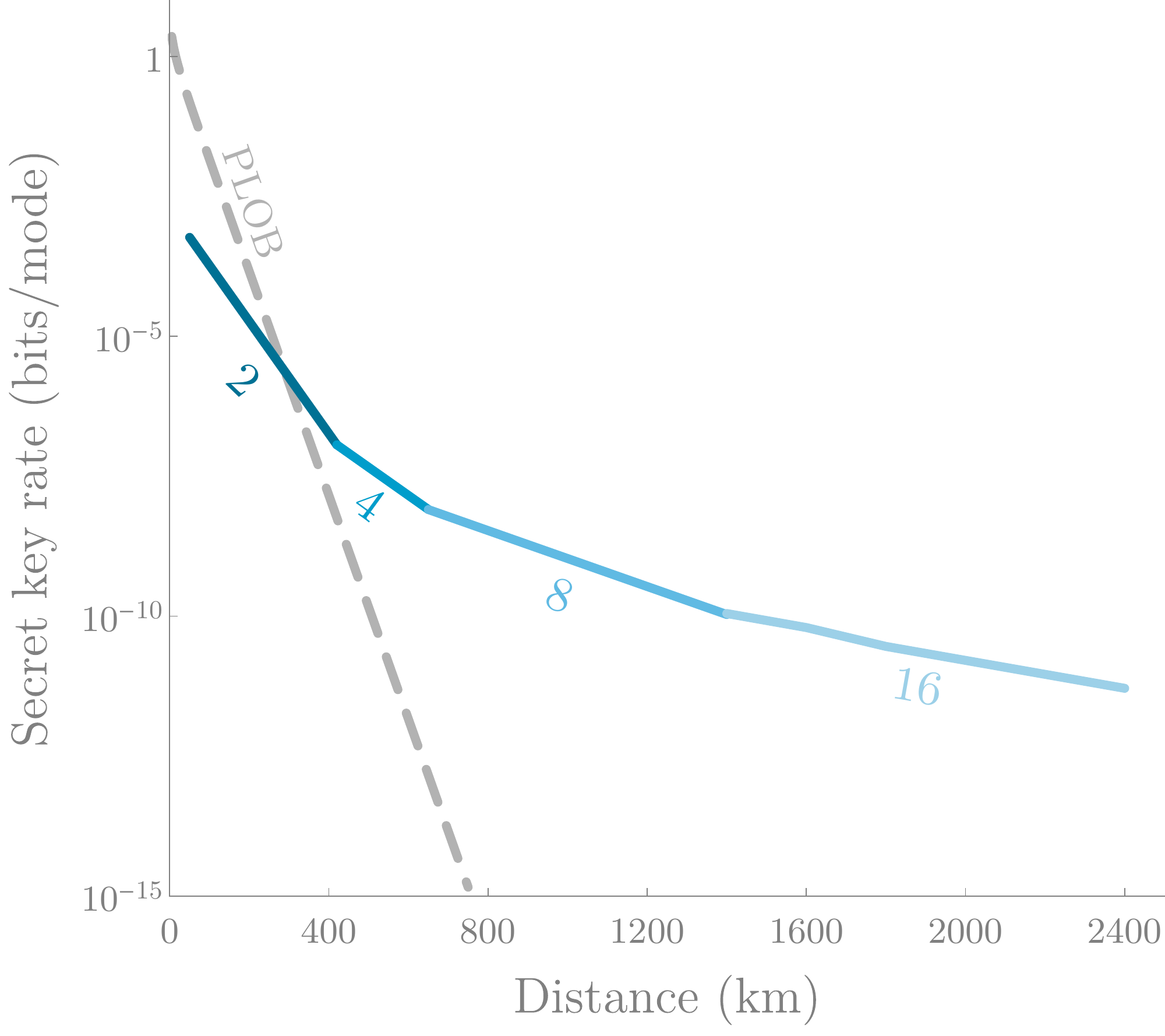}
\caption{Upper bound on key rates of the CV quantum repeater. The solid, blue lines represent different numbers of repeater nodes (repeater links) used along the channel, each line is labeled with the number of links. Post-selection cut-off has been set to \(\gamma_{\mrm{max}}=0.5\). The CV repeater rates shown are for a homodyne-based CVQKD protocol and assume a reconciliation efficiency of \(95\%\). The dashed, light gray line is the PLOB bound \cite{pirandola2017fundamental}.}
\label{fig:PS1}
\end{figure}

We then consider how CV repeater performance scales with distance in the use of CVQKD. In Fig. \ref{fig:PS1}, we give an upper bound on the secret key rate of our CV quantum repeater and show how it changes with more repeater nodes. Determining the actual output state of the multi-node CV repeater is intractable  because it involves integrating over all dual homodyne outcomes $\gamma$. However, an upper bound on the raw secret key rate \(K\) can be calculated from the ideal (\(\gamma=0\)) case, multiplied by the realistic rate of successful operation of the entire repeater \(R_{\mrm{rep}}\), determined numerically. In the case of two links, where the upper bound and the exact numerical result can be compared, we find the two results are close (see Appendix~\ref{app:upper}). Given the favorable performance of our repeater with the homdyne-based CVQKD protocol in Fig.~\ref{fig:result1}, we present results in Fig.~\ref{fig:PS1} focusing only on performance with the homodyne-based protocol. The repeater rate \(R_{\mrm{rep}}\) was obtained via \eqref{eq:reprate} with post-selection probabilties calculated numerically and the post-selection cutoff is fixed at all instances to \(\gamma_{\mrm{max}}=0.5\) (see details in Appendix~\ref{app:multi}). For smaller post-selection cutoffs, the output state yields a higher raw key rate due to higher correlations, however it comes at the expense of a lower probability of successful post-selection.  In Appendix~\ref{app:lower}, we also provide a lower bound on the secret key rate of the CV quantum repeater. 

\section{Conclusion \label{sec:conc}}
In summary, we have presented here a novel scheme for a CV repeater. We emphasize our approach here is different to that of Ref.~\cite{dias2017quantum} as we focus on distribution of CV entanglement, rather than preparing an improved channel. We have shown here that even with reasonably small NLA gains \(g\leq6\), our repeater can distribute entanglement hundreds of kilometres beyond what is achievable with an unphysical, infinitely squeezed TMSV state via direct transmission.  Additionally, we have shown that when these distributed entangled states are used for CVQKD, we are able to improve upon the rates achieved from a previous CV repeater in the literature \cite{dias2017quantum}. In our view, this improvement is attributed to the use of the optimal Gaussian entanglement swapping protocol described in Ref.~\cite{hoelscherobermaier2011optimal} in conjunction with post-selection. Despite the entanglement swapping being non-deterministic due to the use of post-selection, we have found here that we can indeed achieve an improvement.

While our CVQKD analysis incorporates non-ideal reconciliation efficiency, it is idealized in all other senses. A remaining question to be answered would be how the performance of our CV repeater is affected by experimental inefficiencies including inefficient single photon sources in the NLAs,  inefficient homodyne detection and imperfect quantum memories. Specifically with inefficient single photon sources, prior work has shown that this inefficiency causes a gain saturation effect thus limiting the actual achievable gain of the NLA \cite{kocsis2012heralded,xiang2010heralded} with maximum reported gains of \(g^2=11\pm 1\) \cite{ulanov2015undoing} . For distances larger than 130km, the gain for optimal operation of our CV repeater is greater than this maximum reported gain. Further improvements in photon production and detection efficiency will be needed to obtain these higher gains, however we note that single photon source efficiency is constantly improving. It is possible that operation of this repeater may be further optimized by use of a different distillation protocol. Consideration of how this scheme performs with different distillation protocols is left for future work. 

\section{Acknowledgments}
We thank William J. Munro for useful discussions. This research was supported by the Australian Government Department of Defence and by the Australian Research Council (ARC) under the Centre of Excellence for Quantum Computation and Communication Technology (Project No.
CE170100012). 

\appendix
\section{Single-node repeater \label{app:single}}
In this section, we outline how to calculate the entangled output state of the single-node repeater protocol (Fig.~2 in the main text). Initial entanglement distribution begins with generating two independent two-mode Gaussian squeezed vacuum states of form: 
\begin{equation}
\ket{\chi}_{AC}= \sqrt{1-\chi^2} \sum_{n=0}^\infty \chi^n \ket{n}_A\ket{n}_C
\end{equation}
The sources are placed with Alice and the repeater node, with one mode of each entangled state distributed to the repeater node and Bob respectively. This is modeled by a pure-loss channel of transmission \(\eta\).  This transforms mode \(C\) as: 
\begin{align}
\begin{split}
\hat{U}_{BS} & \left[ \ket{n}_C\ket{0}_D \right] 
\\
& = \sum_{p =0}^n \sqrt{\binom{n}{p}} \eta^{p/2} \left(1-\eta\right)^{(n-p)/2} \ket{p}_C\ket{n-p}_D 
\end{split}
\end{align}
where mode \(D\) is an environment mode. The state becomes:
\begin{align}
\begin{split}
\ket{\chi}_{AC}  \to  & \sqrt{1-\chi^2} \sum_{n=0}^\infty \sum_{p =0}^n  \chi^n \ket{n}_A
\\
& \sqrt{\binom{n}{p}} \eta^{p/2} \left(1-\eta\right)^{(n-p)/2} \ket{p}_C \ket{n-p}_D 
\end{split}
\end{align}
Entanglement distillation proceeds by acting an NLA on mode \(C\) with gain \(g\). The action of the NLA with a single quantum scissor can be described by the following operation \cite{dias2017quantum}:
\begin{equation}
\hat{T}_1 = \hat{\Pi}_1 g^{\hat{n}}
\end{equation}
where the truncation operator $\hat{\Pi}_1$ is defined as:
\begin{equation}
\hat{\Pi}_1 = {\frac{1}{\sqrt{g^2+1}}} \left(\ket{0}\bra{0}+\ket{1}\bra{1}\right)
\end{equation}
After this operation, the $\ket{2}$ and higher order photon terms in mode $C$ are truncated and the state becomes: 
\begin{align}
\begin{split}
&\ket{\psi}_{ACD}=  \sqrt{\frac{1-\chi^2}{g^2+1}} 
\left(  \sum_{n=0}^\infty  \chi^n \left(1-\eta\right)^{n/2} \ket{0}_C \ket{n}_A   \ket{n}_D  
\right. 
\\
& \left.+ g  \sqrt{\eta} \sum_{n=1}^\infty  \chi^n  \sqrt{n} \left(1-\eta\right)^{(n-1)/2} \ket{1}_C \ket{n}_A  \ket{n-1}_D \right)
\end{split}
\label{eq:EPRlossQS}
\end{align}
The probability of success of this individual NLA can be found via the norm of the un-normalised state \eqref{eq:EPRlossQS} which is:
\begin{equation}
P_{NLA} =  \frac{\left(1-\chi ^2\right) \left(\chi ^2 \left(\eta  g^2+\eta -1\right)+1\right)}{\left(g^2+1\right) \left((\eta -1) \chi ^2+1\right)^2}
\label{eq:Psucc}
\end{equation}
The final step in this single-node repeater protocol is the entanglement swapping operation. We use a second copy of the state \eqref{eq:EPRlossQS}, with modes \(F\) and \(B\)  distributed between the repeater node and Bob respectively, given by:
\begin{align}
\begin{split}
& \ket{\psi}_{BFE}= \sqrt{\frac{1-\chi^2}{g^2+1}} 
\left(  \sum_{m=0}^\infty  \chi^m \left(1-\eta\right)^{m/2} \ket{0}_B  \ket{m}_F   \ket{m}_E \right.
\\
& \left. + g  \sqrt{\eta} \sum_{m=1}^\infty  \chi^m  \sqrt{m} \left(1-\eta\right)^{(m-1)/2} \ket{1}_B \ket{m}_F   \ket{m-1}_E \right)
\label{eq:EPRlossQS2}
\end{split}
\end{align}
where mode $E$ is an environment mode. With these two entangled states \eqref{eq:EPRlossQS} and \eqref{eq:EPRlossQS2}, modes $F$ and $C$ are combined at the repeater node and a dual homodyne detection is performed. To model this dual homodyne detection,  we project modes \(F\) and \(C\) onto the eigenstate \cite{hofmann2000fidelity, ide2001continuous}: 
\begin{equation}
 \ket{\gamma}_{FC} = \frac{1}{\sqrt{\pi}}\sum_{n=0}^\infty \hat{D}_C(\gamma)\ket{n}_C\ket{n}_F
\label{eq:proj}
 \end{equation}
where \(\gamma\) corresponds to the measurement outcome of the dual homodyne detection. The state after swapping can be found via:
\begin{equation}
\ket{\psi_{\mathrm{swap}}}_{ABDE}= \bra{\gamma}_{FC} \left[\ket{\psi}_{ACD}\otimes \ket{\psi}_{BFE} \right]
\label{eq:symSwap} 
\end{equation}
From Eq.~\eqref{eq:symSwap} after corrective displacements on modes \(A\) and \(B\) , we find the following un-normalized entangled state shared between Alice and Bob (including environment modes $D$ and $E$) and conditioned on the measurement outcome of $\gamma$:
\begin{widetext}
\begin{align}
\begin{split}
&\ket{\psi_{out}}_{ABDE}= \frac{1}{\sqrt{\pi}}  \frac{1-\chi^2}{g^2+1}  e^{-\left|\gamma\right|^2/2}  \hat{D}_A\left(\lambda_a \gamma\right) \hat{D}_B \left(\lambda_b \gamma\right)
\\
&   \bigg[  \sum_{n=0}^\infty \sum_{m=0}^\infty \chi^n \left(1-\eta\right)^{n/2}   \chi^m \left(1-\eta\right)^{m/2}  \frac{\left(-\gamma\right)^m}{\sqrt{m!}}  \ket{n}_D   \ket{n}_A    \ket{0}_B \ket{m}_E 
\\
&  +  g  \sqrt{\eta}  \sum_{n=0}^\infty  \sum_{m=1}^\infty   \chi^n \left(1-\eta\right)^{n/2}   \chi^m  \sqrt{m} \left(1-\eta\right)^{(m-1)/2}  \frac{\left(-\gamma\right)^m}{\sqrt{m!}}   \ket{n}_D \ket{n}_A \ket{1}_B \ket{m-1}_E
\\
&  + g  \sqrt{\eta} \sum_{n=1}^\infty    \sum_{m=0}^\infty  \chi^n  \sqrt{n} \left(1-\eta\right)^{(n-1)/2}  \chi^m \left(1-\eta\right)^{m/2} \gamma^* \frac{\left(-\gamma\right)^m}{\sqrt{m!}}   \ket{n-1}_D  \ket{n}_A  \ket{0}_B \ket{m}_E 
\\
&  + g  \sqrt{\eta} \sum_{n=1}^\infty    \sum_{m=1}^\infty  \chi^n  \sqrt{n} \left(1-\eta\right)^{(n-1)/2}  \chi^m \left(1-\eta\right)^{m/2}  \sqrt{m}\frac{\left(-\gamma\right)^{m-1}}{\sqrt{\left(m-1\right)!}}  \ket{n-1}_D  \ket{n}_A  \ket{0}_B \ket{m}_E 
\\
&  + g^2  \eta \sum_{n=1}^\infty \sum_{m=1}^\infty  \chi^n  \sqrt{n} \left(1-\eta\right)^{(n-1)/2}   \chi^m  \sqrt{m} \left(1-\eta\right)^{(m-1)/2}  \gamma^* \frac{\left(-\gamma\right)^m}{\sqrt{m!}}\ket{n-1}_D \ket{n}_A \ket{1}_B \ket{m-1}_E
\\
&  + g^2  \eta \sum_{n=1}^\infty \sum_{m=1}^\infty  \chi^n  \sqrt{n} \left(1-\eta\right)^{(n-1)/2}   \chi^m  \sqrt{m} \left(1-\eta\right)^{(m-1)/2}   \sqrt{m} \frac{\left(-\gamma\right)^{m-1}}{\sqrt{\left(m-1\right)!}} \ket{n-1}_D \ket{n}_A \ket{1}_B \ket{m-1}_E
  \bigg]
  \label{eq:rhoAll}
\end{split}
\end{align}
\end{widetext}
where \(\lambda_a\) and \(\lambda_b\) correspond to the classical gains applied to scale the displacements on modes \(A\) and \(B\) respectively. The density matrix of the output state shared between Alice and Bob can be found via:
\begin{equation}
\hat{\rho}_{AB} \left(\gamma \right)= \mathrm{Tr}_{DE}\left[ \ket{\psi_{out}}_{ABDE}\bra{\psi_{out}}_{ABDE}\right]
  \label{eq:rhoOut}
\end{equation}

To find the probability of successful post-selection \(P_{\mrm{PS}}\), we use the following:
\begin{equation}
P_{\mrm{PS}} = \frac{\int_0^{2\pi} \int^{\gamma_{\mrm{max}}}_{0} \mathrm{Tr} \hat{\rho}_{AB} \left(\gamma\right)|\gamma| \, \mathrm{d}\phi_\gamma  \mrm{d}|\gamma|}{\int_0^{2\pi} \int_{0}^\infty \mathrm{Tr} \hat{\rho}_{AB} \left(\gamma\right)|\gamma|\,\mrm{d}\phi_{\gamma} \mathrm{d} |\gamma| }
\label{eq:pps}
\end{equation} 

From the entangled output state \eqref{eq:rhoOut} shared between Alice and Bob, we are now in a position to calculate the secret key rate assuming collective attacks given by \cite{garcia2007quantum}:
\begin{equation}
K= \beta I_{AB} - I_{E}
\label{eq:skr}
\end{equation}
where $I_{AB}$ is the mutual information shared between Alice and Bob, $I_E$ is the Holevo bound representing the maximum amount of quantum information accessed by Eve, and $0\leq \beta\leq 1$ is the reconciliation efficiency.  
We calculate the key rate from the covariance matrix of the entangled  output state shared between Alice and Bob. The covariance matrix elements were obtained using \eqref{eq:rhoAll} and \eqref{eq:rhoOut} and averaged over the accepted post-selection region. To be more specific, we accept results \(\gamma\) of the dual HD which fall in a circular region centered on the origin to some maximum radius \(\gamma_{\mrm{max}}\). Averaging was performed via numerical integration of each covariance matrix element.
A two-mode Gaussian state has covariance matrix in standard form:
\begin{equation}
V = \begin{bmatrix}
a\mathds{1}  &  c \mathbb{Z} \\
c \mathbb{Z} & b \mathds{1}
\end{bmatrix} .
\end{equation}
Even though the output entangled state is slightly non-Gaussian due to the the QS operation and thus cannot be fully characterized by its covariance matrix, it is valid to use Gaussian key rate calculations as it overestimates Eve's information \cite{wolf2006extremality,navascues2006optimality,garcia-patron2006unconditional}. We calculate the mutual information shared between Alice and Bob $I_{AB}$ for an entanglement-based protocol where Alice and Bob both conduct heterodyne detection on their entangled modes by \cite{weedbrook2004quantum,garcia2007quantum}:
\begin{equation}
I_{AB}^{\mrm{het}} = \log_2 \left( \frac{1+a}{1+a- \frac{c^2}{1+b}} \right) 
\end{equation}
and for an entanglement based protocol where Alice and Bob conduct homodyne detection \cite{cerf2001quantum, garcia2007quantum}:
\begin{equation}
I_{AB}^{\mrm{hom}} =\frac{1}{2} \log_2 \left( \frac{a}{a-\frac{c^2}{b}} \right) 
\end{equation}
We illustrate here how we calculate Eve's information given Bob as the reference for reconciliation which is the case for reverse reconciliation, giving $I_E=I_{BE}$ and representing the mutual information between Bob and Eve (for direct reconciliation where Alice is the reference we would have $I_E=I_{AE}$).  Eve's information $I_{BE}$ can be calculated via:
\begin{equation}
I_{BE} = S(E) -S(E|B) 
\label{eq:IEB}
\end{equation}
 where $S(E)$ is the Von-Neumann entropy of Eve's state before measurement and $S(E|B)$ is the Von-Neumann entropy of Eve's state conditioned on Bob's measurement outcome. $S(E)$ can be found by using the fact that Eve purifies Alice and Bob's system, giving $S(E) =S(AB)$ which is defined as:
 \begin{equation}
 S(AB) = G\left( \frac{\nu_1-1}{2} \right) +G\left(\frac{\nu_2-1}{2} \right)
 \end{equation}
 where $\nu_1$ and $\nu_2$ are the symplectic eigenvalues of the covariance matrix $V$ and 
 \begin{equation}
 G(x) = \left(1+x\right) \log_2 \left(1+x\right)- x \log_2 x . 
\end{equation}  
The symplectic eigenvalues $\nu_1$ and $\nu_2$ can be found via:
\begin{equation}
\nu_{1,2} = \sqrt{\frac{\Delta\pm \sqrt{\Delta^2-4 \det V}}{2}}
\end{equation}
where $\Delta= a^2+b^2 -2c^2$. The Von-Neumann entropy of the conditional state $S(E|B)$ is a function of the symplectic eigenvalue of the conditional covariance matrix, $\nu_3=a-\frac{c^2}{1+b}$:
\begin{equation}
S(E|B) = G \left(\frac{\nu_3-1}{2} \right) .
\end{equation}

\section{Multi-node repeater\label{app:multi}}
To go beyond the simplest case of two links we proceed by using two copies of the state \eqref{eq:rhoOut} which have been distributed along four initial segments of the channel:
\begin{equation}
\hat{\rho}_{ABMN} \left(\gamma_1, \gamma_2\right) = \hat{\rho}_{AB} \left(\gamma_1\right) \otimes  \hat{\rho}_{MN} \left(\gamma_2\right)
\label{eq:abmn}
\end{equation}
where both \(\hat{\rho}_{AB} \left(\gamma_1\right) \) and \(\hat{\rho}_{MN} \left(\gamma_2\right)\) correspond to the output state \eqref{eq:rhoOut} conditioned on measurement outcomes \(\gamma_1\) and \(\gamma_2\) from dual HDs at nodes 1 and 3 respectively (see Fig.~\ref{fig:nestedAsym}).  Modes $B$ and $M$ are mixed at the central node and a dual HD is conducted on both modes, giving the total conditional output state:
\begin{align}
\begin{split}
\hat{\rho}_{AN}&\left(\gamma_1, \gamma_2, \gamma_3 \right) 
\\
&= \mathrm{Tr}_{BM} \left[\hat{\rho}_{ABMN} \left(\gamma_1, \gamma_2\right) \otimes \ket{\gamma_3}_{BM}\tensor[_{BM}]{\bra{\gamma_3}}{}  \right]
\label{eq:an}
\end{split}
\end{align}
Finally, the output modes $A$ and $N$ are displaced by the measurement outcome $\gamma_3$ scaled by classical gains \(\lambda_a\) on mode \(A\) and \(\lambda_n\) on mode \(N\): 
\begin{align}
\begin{split}
&\hat{\rho}_{out} \left(\gamma_1, \gamma_2, \gamma_3 \right)  
\\
&=\hat{D}_N \left( \lambda_n \gamma_3 \right) \hat{D}_A \left( \lambda_a \gamma_3 \right) \hat{\rho}_{AN}\left(\gamma_1, \gamma_2, \gamma_3 \right)
\\
& \hspace{20pt}\hat{D}_N^\dagger \left( \lambda_n \gamma_3 \right) \hat{D}_A^\dagger \left( \lambda_a \gamma_3 \right)
\end{split}
\end{align}
We have outlined here the process for calculating the output state of four links of the CV quantum repeater, the output state of eight and higher links proceeds in the same way. 
\subsection{Lower bound\label{app:lower}}
Evaluating performance of our CV repeater via the method outlined in the previous section is not tractable for the multi-node repeater. This is because results obtained need to be integrated over each dual HD measurement outcome. While for two links, results can be obtained via numerical integration (and are given in Fig.~\ref{fig:result1}), for four and higher links we will model performance of our CV repeater by averaging the output density matrix after each entanglement swapping step.
That is, instead of \eqref{eq:abmn}, the density matrices are first averaged over the accepted post-selection region: 
\begin{equation}
\hat{\rho}_{AB}\to \int_0^{2\pi} \int^{\gamma_{\mrm{max}}}_{0} \hat{\rho}_{AB} \left(\gamma\right) |\gamma|\,\mathrm{d} \phi_{\gamma} \mrm{d} |\gamma|
\label{eq:avrho}
\end{equation}
As previously described, in the four link repeater scheme, two copies of the  state are used:
\begin{equation}
\hat{\rho}_{ABMN} = \hat{\rho}_{AB} \otimes  \hat{\rho}_{MN} 
\end{equation}
The two averaged output states are then combined and swapped:
\begin{align}
\begin{split}
\hat{\rho}_{AN}&\left(\gamma_3 \right) 
\\
&= \mathrm{Tr}_{BM} \left[\hat{\rho}_{ABMN} \otimes \ket{\gamma_3}_{BM}\tensor[_{BM}]{\bra{\gamma_3}}{}  \right]
\end{split}
\end{align}
This is followed by a displacement on modes \(A\) and \(N\):
\begin{align}
\begin{split}
&\hat{\rho}_{out}\left(\gamma_3 \right)  
\\
& \hspace{15pt}= \hat{D}_N \left( \lambda_n \gamma_3 \right) \hat{D}_A \left( \lambda_a \gamma_3 \right) \hat{\rho}_{AN}\left(\gamma_3 \right) \hat{D}_N^\dagger \left( \lambda_n \gamma_3 \right) \hat{D}_A^\dagger \left( \lambda_a \gamma_3 \right)
\label{eq:avgout}
\end{split}
\end{align}
 The success probability of the final entanglement swap at the central node is given by:
\begin{equation}
P_{PS} = \frac{\int_0^{2\pi} \int^{\gamma_{\mrm{max}}}_{0} \mathrm{Tr}\hat{\rho}_{AN} \left(\gamma_3 \right)|\gamma_3| \, \mathrm{d} \phi_{\gamma_3} \mrm{d} |\gamma_3| }{\int_0^{2\pi} \int^{\infty}_{0}\mathrm{Tr} \hat{\rho}_{AN}\left(\gamma_3 \right)  |\gamma_3|\,\mathrm{d} \phi_{\gamma_3} \mrm{d} |\gamma_3|}
\label{eq:Pnest}
\end{equation}
Calculation of the covariance matrix proceeds by using the output state \eqref{eq:avgout} and numerically integrating to average over the accepted post-selection region. 

By averaging the density matrices before input into subsequent entanglement swapping, the calculations become tractable. However, as \(\gamma\) is a classical parameter, this averaging will unavoidably lead to an overestimation in the noise present in the output state. Therefore, we present the results gained from this method as a lower bound to the key rates achievable by our CV quantum repeater. While this method may be used to estimate the probability of success of the nested entanglement swapping operations \eqref{eq:Pnest} and thus can be used to estimate \(R_{\mathrm{rep}}\), the raw key rate \(K\) calculated from the covariance matrix of the output state \eqref{eq:avgout} will suffer from the overestimation of noise. 

For four links of the CV repeater, the repeater rate \eqref{eq:reprate} is given by:
\begin{equation}
R_{\mrm{rep}}  =  \frac{1}{Z_2\left(P_{\mrm{NLA}}\right) } \times  \frac{1}{Z_1\left(P_{\mrm{PS}1}\right)} \times  \frac{1}{Z_0\left(P_{\mrm{PS}0}\right) }
\end{equation}
where \( P_{\mrm{PS1}}\) is the probability of successful post-selection in the 2 base level entanglement swaps, and \(P_{\mrm{PS}0}\) is the probability of success of post-selection in the single higher level entanglement swap.

\subsection{Upper bound\label{app:upper}}
An upper bound on the raw secret key rate \(K\) can be determined from the ideal output state of the CV repeater protocol. That is, the output state achieved conditioned on the measurement outcome of \(\gamma=0\), which results in no displacement. In this ideal case, the covariance matrices for the output states of the two, four, eight and sixteen link schemes are analytically solvable. We use the raw key rate \(K\) calculated from the ideal (\(\gamma=0\)) case, multiplied by the realistic rate of successful operation of the entire repeater \(R_{\mrm{rep}}\). This rate depends on the success probability of the NLA \eqref{eq:Psucc}, and the probabilities of success of post-selection calculated via the method explained in the previous section \eqref{eq:pps} and  \eqref{eq:Pnest}. It is through this method that the results in Fig.~\ref{fig:PS1} were obtained. 
\begin{figure}
\centering
\includegraphics[width=0.99\linewidth]{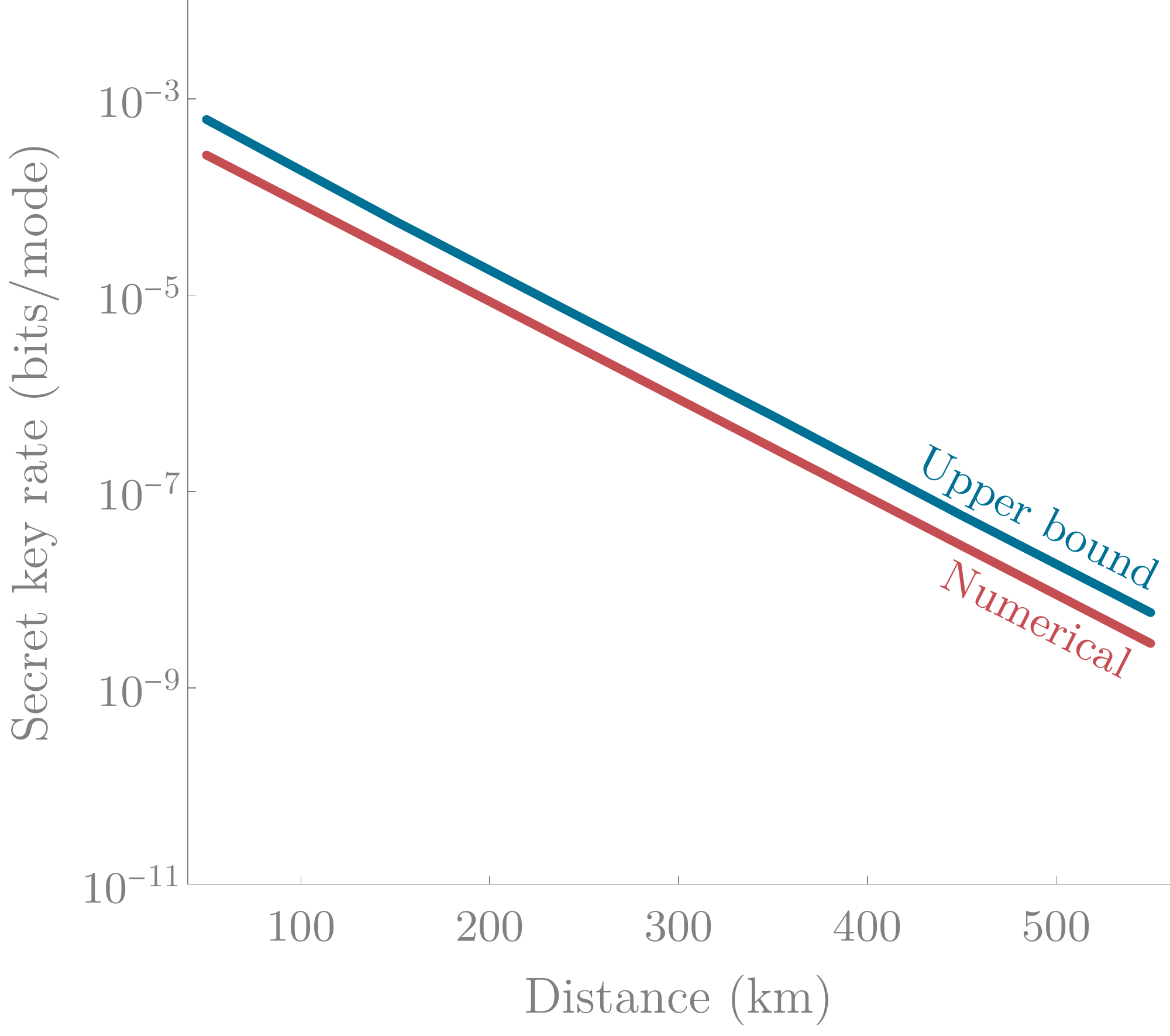}
\caption{Comparison between upper bound on the secret key rate using an ideal covariance matrix and that using numerically integrated output state \eqref{eq:rhoOut}. The dark blue line corresponds to the two-link upper bound shown in Fig.~\ref{fig:PS1}, while the light blue line corresponds to the two link secret key rate shown in  Fig.~\ref{fig:result1}.}
\label{fig:comp}
\end{figure}

\begin{figure}
\centering
\includegraphics[width=0.99\linewidth]{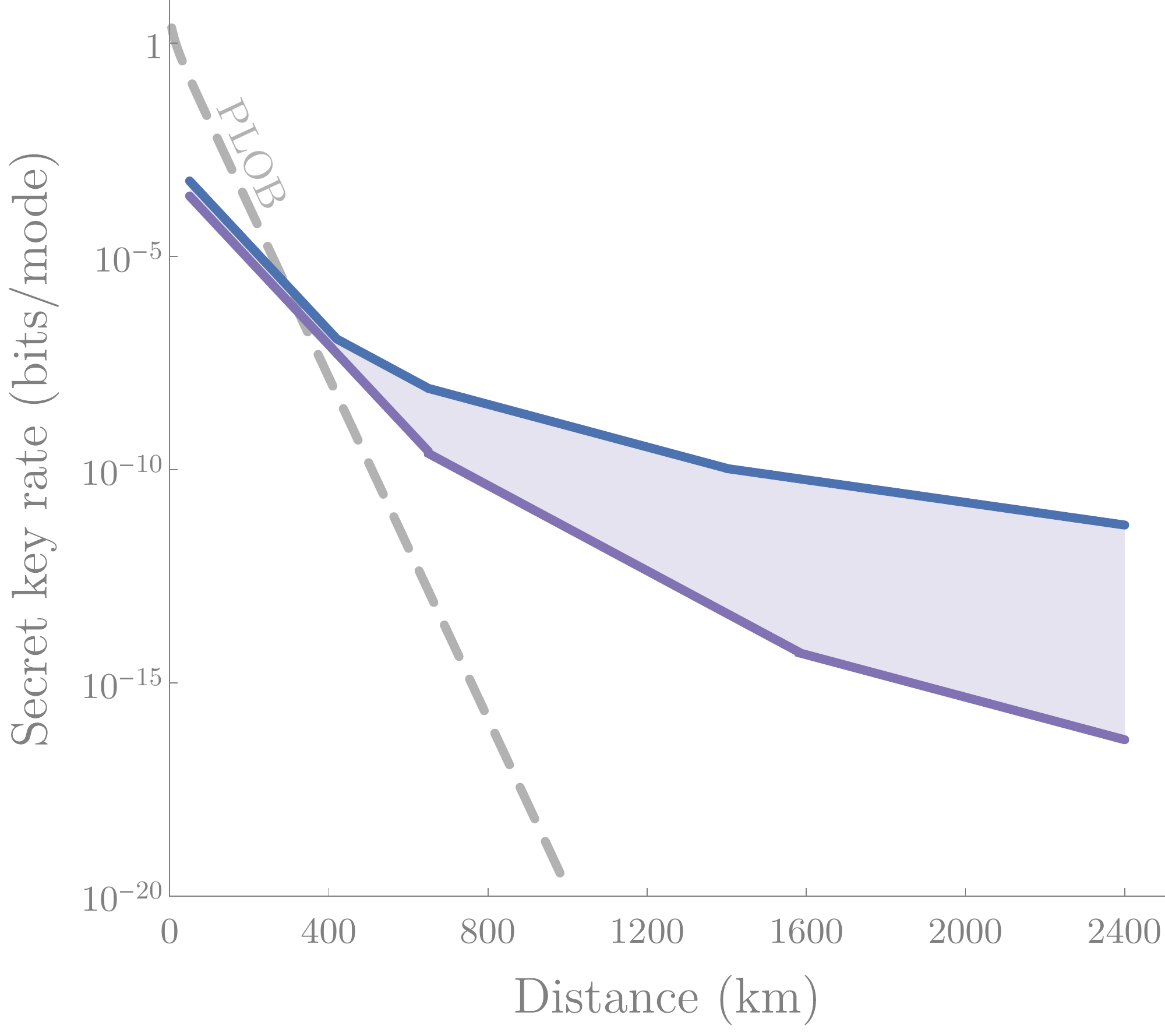}
\caption{Lower and upper bounds on key rates of the CV quantum repeater. The solid blue line corresponds to the upper bound shown in Fig.~\ref{fig:PS1}. The solid purple line corresponds to the lower bound calculated via the method in Appendix \ref{app:lower}.}
\label{fig:krL}
\end{figure}
We can compare the upper bound to the secret key rate of the post-selected output state in the two link case and this is shown in Fig.~\ref{fig:comp}. As can be seen in Fig.~\ref{fig:comp}, the upper bound and numerically integrated key rates are quite close. This is because the covariance matrices of the ideal output state and the post-selected output state for \(\gamma_{\mrm{max}}=0.5\) are close.

Finally, we can use our upper bound to give the region of estimated performance of our CV quantum repeater and this is shown in Fig.~\ref{fig:krL}. As previously noted, the upper bound uses the fixed post-selection cut-off of \(\gamma_{\mrm{max}}=0.5\) at all swapping levels. However, the post-selection cut-off of the lower bound varies at each level. This is because calculating the lower bound requires small post-selection cut-offs for initial swaps (i.e. \(\gamma\) close to 0) since we swap average density matrices \eqref{eq:avrho}). Rough optimisation of the overall key rates including raw key rate and repeater rate yields the following post-selection cut-offs.  For the results in Fig.~\ref{fig:krL}, the two link lower bound uses a cut-off of \(\gamma_{\mrm{max}}=0.5\). The four-link lower bound uses post-selection cut-off of \(\gamma_{\mrm{max}}=0.2\) at the base level and \(\gamma_{\mrm{max}}=0.45\) at the upper level entanglement swap. Lastly, the eight-link lower bound uses cut-offs  \(\gamma_{\mrm{max}}=0.06,\,0.15,\,0.4\) at the base, mid and highest level entanglement swaps respectively. Note that the region between upper and lower bound increases for longer distances due to the compounding effect of noise from averaging after multiple entanglement swapping rounds as well as the reduction in lower bound repeater rate due to the smaller post-selection cut-offs.

\end{document}